\documentstyle[epsfig]{mn}

\input{psfig.sty}

\ifnfsstwo

\fi

\ifnfssone
  \newmathalphabet{\mathit}
    \addtoversion{normal}{\mathit}{cmr}{m}{it}
    \addtoversion{bold}{\mathit}{cmr}{bx}{it}

\fi

\ifoldfss

\fi

\loadboldmathitalic
\loadboldgreek

\title[A new formulation of the Type Ia SN rate]
       {A new formulation of the Type Ia SN rate and its consequences on 
galactic chemical evolution}
\author[F. Matteucci et al.]
 {F. Matteucci$^{1,2}$, N. Panagia$^{3}$, A. Pipino$^{1}$, 
F. Mannucci$^{4}$, S. Recchi$^{5,2}$, and M. Della Valle$^{6}$ \\
$^1$ Dipartimento di Astronomia, Universita' di Trieste,
 Via G.B. Tiepolo 11, 34131 Trieste, Italy \\
 $^2$ INAF, Osservatorio Astronomico di Trieste,  Via G. B. Tiepolo 11, 34131 
Trieste, Italy\\
$^{3}$ StSci, 3700 san Martin Drive, Baltimore, MD 21218, USA\\
$^{4}$ INAF-IRA, Largo E. Fermi 5, 50125 Firenze, Italy\\
$^{5}$ Institute of Astronomy, Tuerkenstrasse 17, 1180 Vienna, Austria\\
$^{6}$ INAF, Osservatorio Astrofisico di Arcetri, Largo E. Fermi 5, 50125 
Firenze, Italy\\}

\pubyear{2006}

\begin{document}
\maketitle

\begin{abstract}
In recent papers Mannucci et al. (2005, 2006) suggested, on the basis of observational arguments, that there is a bimodal distribution of delay times for the explosion of Type Ia SNe. In particular, a percentage from 35 to 50\% of 
the total Type Ia SNe should be composed by systems with lifetimes as 
short as $10^{8}$ years, whereas the rest should arise from smaller mass progenitors with a much broader distribution of lifetimes. 
In this paper, we test this hypothesis in models of chemical evolution of 
galaxies of different morphological 
type: ellipticals, 
spirals and irregulars. We show that this proposed scenario is compatible also 
with the main chemical properties of galaxies.
In this new formulation, we simply assume that Type Ia SNe are originating 
from C-O white dwarfs in binary systems without specifying if the progenitor 
model is the single-degenerate or the
double degenerate one or a mixture of both. In the framework of the single 
degenerate model, such a  bimodal distribution of the time delays could be 
explained if the binary systems with a unitary mass ratio are 
favored in the mass range 5-8$M_{\odot}$, 
whereas for masses $< 5M_{\odot}$  the favored systems should have the mass 
of the primary much larger than the mass of the secondary.
When the new rate is introduced in the two-infall model for the Milky Way, 
the derived Type Ia SN rate as a function of cosmic 
time shows a high and broad peak at very early 
epochs thus influencing the chemical evolution of the galactic halo more 
than in the previous widely adopted 
formulations for the SNIa rate (for example Greggio \& Renzini, 1983). 
As a consequence of this, the [O/Fe] ratio decreases faster 
for [Fe/H] $>$ -2.0 dex, relative to the old models. For a typical 
elliptical of $10^{11} M_{\odot}$ of luminous mass, the new rate produces
average [$\alpha$/ Fe] ratios in the dominant stellar population still 
in agreement with observations.
The Type Ia SN rate also in this case shows an earlier peak and a 
subsequent faster decline relative to the previous results, but the differences are smaller than 
in the case of our Galaxy. 
We have also checked the effects of the new Type Ia SN rate on  the evolution 
of the Fe content in the ICM, as a consequence of its production from cluster 
ellipticals and we found that less Fe in the ICM is produced with the new rate, due to the higher fraction of Fe synthesized at early times and remaining locked into the stars in ellipticals. 
For dwarf irregular galaxies suffering few bursts of 
star formation we obtain [O/Fe] ratios larger by 0.2 dex relative to the 
previous models.

\end{abstract}

\begin{keywords}
galaxies: abundances  --
galaxies: evolution -- supernovae: general
\end{keywords}

\section{Introduction}
The Type Ia supernova (SN) rate is a fundamental ingredient in models of 
galactic evolution.
In the pioneering work of Greggio \& Renzini (1983a, hereafter GR83) 
there was, 
for the first time, an expression for the Type Ia SN rate in the scenario of 
the single degenerate model. 
In this scenario, SNe Type Ia arise from the explosion of a C-O white dwarf in 
a close binary system where the companion is either a red giant or a main 
sequence star (Whelan \& Iben, 1973; Munari \& Renzini, 1992; 
Kenyon et al.1993; Hachisu et al., 1996; 1999).
An alternative model for progenitors of Type Ia SNe was proposed by 
Iben \& Tutukov (1984). In this scenario two C-O white dwarfs of $\sim 0.7 
M_{\odot}$ merge after loss of angular momentun due to gravitational wave 
emission, and explode
since the final object reaches the Chandrasekhar mass. Tornamb\'e \& Matteucci 
(1986) formulated a Type Ia SN rate in this scenario and applied it to 
galactic chemical evolution models.
In the following,  Matteucci \& Greggio (1986) tested the GR83 rate by means of a detailed chemical evolution of the Milky Way and interpreted the [$\alpha$/Fe] ratios versus [Fe/H] as due to the delay in the Fe production from Type
Ia SNe, thus confirming previous suggestions (Tinsley, 1979; Greggio \& Renzini 1983b). They suggested that the time scale for the change in the slope of this 
relation, due to the Fe restored in a sustantial amount by Type  Ia SNe is 
$\sim 1-1.5$ Gyr.
This timescale is not universal but related  to the specific history of 
star formation in galaxies, as shown by Matteucci \& Recchi (2001, hereafter 
MR01), being shorter than in the solar neighbourhood for ellipticals and 
longer for irregular systems.
The formulation of the SNIa rate in the single degenerate scenario was later 
implemented by Greggio (1996), who considered also the possibility of 
sub-Chandrasekhar masses.  
In the single degenerate scenario the clock for the esplosion is given by the 
lifetime of the secondary mass which transfers material over the primary star 
when it fills its Roche lobe. In the original formulation of GR83 and in 
the one of MR01 the mass range 
for the progenitors of both the primary and secondary masses 
is 0.8-8$M_{\odot}$. 
The upper limit is given by the fact that stars with masses $M> 8 M_{\odot}$ 
ignite carbon in a non degenerate core and therefore do not end their lives as 
C-O white dwarfs. The lower limit is instead obviously due to the fact that 
we are only interested in systems which can produce a Type Ia SN in a Hubble 
time.
In this scenario the very first binary system exploding as a Type Ia SN is 
made of two stars of $8M_{\odot}$ and therefore it occurs only 30-40 Myr since 
the beginning of star formation.
\par
Recently, Mannucci et al. (2005) showed that two populations of 
progenitors of Type Ia SNe are needed to explain the dependence of the rates 
on the colors of the parent galaxy. The presence of Type Ia SNe in old, red, 
quiescent galaxies is an indication that part of these SNe originate from old 
stellar populations. On the contrary, the increase of the rate in blue galaxies
(by a factor of 30 going from (B-K) $\sim$2 to (B-K) $\sim$ 4.5) shows that 
part of the SN Ia progenitors is related to young stars and closely follows 
the evolution of the  star formation rate.
Mannucci, Della Valle \& Panagia (2006, hereafter MVP06), on the 
basis of the previously described relation between the Type Ia SN rate and the color of the parent galaxies, their radio power as measured by Della Valle et al. (2005), and cosmic age, 
concluded that there are two populations of progenitors of 
Type Ia SNe. They have demonstrated that the current observations can be accounted for only if about half of the SNe Ia ({\it prompt SNe Ia}) explode within $10^{8}$ years after the formation of their progenitors, while the rest explode during a wide period of time extending up to 10 Gyrs ({\it tardy SNe Ia}). 
As already mentioned, the GR83 rate in its original formulation contains a 
population of Type Ia SNe 
exploding rapidly, at variance with other progenitor models suggested in the 
literature such as that of 
Kobayashi et al. (1998),  where the maximum mass for the secondary stars is 
assumed to be 2.6 $M_{\odot}$ 
and therefore has a delay time distribution function (hereafter DTD), namely the distribution of the explosion times, quite different from the one of GR83. 
In general, no Type Ia SN rate proposed in the literature up to now (GR83, 
Yungelson \& Livio, 2000; 
MR01, Belczynsky et al 2005; Greggio 2005), both in the framework of 
single and 
double-degenerate scenarios, can reproduce the bimodality of the DTDerived by 
MVP06. In fact, although some of these models (GR83 and MR01, 
single degenerate; Greggio 2005, double degenerate) predict  a  
DTD in agreement with the evolution of the 
SNIa rate with redshift and with galaxy color, none of them predicts 
enough SNIa within the first $10^{8}$ 
years to fully account for the radio dependence of the rate in ellipticals.

The aim of this paper is 
to adopt this new DTD in models of 
chemical evolution of galaxies 
(ellipticals, spirals and irregulars).
In particular, we will calculate the Type Ia SN rate in different galaxies by 
adopting different star formation 
rates, then we will compare the chemical evolution results obtained with this 
new rate with the observations and 
with the results of previous models adopting the formulation of the 
Type Ia SN rate of MR01.
Scannapieco \& Bildsten (2005) used the results of Mannucci et al. (2005) 
to derive, in the framework of a simple closed-box model, the evolution of the 
SN Ia rate and the chemical evolution of the Milky Way, as well as the 
evolution of the Fe content in the intracluster medium. 

Here, we will apply the new Type Ia rate formulation to very detailed chemical 
evolution models for galaxies of different morphological type, taking into 
account both infall and outflow and already reproducing the majority of 
observational constraints.

The paper is organized as follows:
in Section 2 the formulations for the SNIa rate of GR83 and MR01 together 
with the new one are presented. In Section 3 the galactic chemical 
evolution models are discussed and in Section 4 the model results are 
shown and compared to each others and with data. Finally,  in Section 
5 some conclusions are drawn.

\section{Type Ia supernovae}
\subsection{Progenitors}

We recall here the most common models for the progenitors of Type Ia
SNe proposed insofar:

\begin{itemize}

\item The merging of two C-O white dwarfs (WDs), due to gravitational wave radiation,
which reach the Chandrasekhar mass and explode by C-deflagration (Iben
and Tutukov 1984). This is known as double-degenerate (DD)
scenario. The progenitor masses should be in the range 5-9$M_{\odot}$ to ensure 
two WDs of $\sim 0.7 M_{\odot}$ in order to reach the Chandrasekhar mass. Moreover the mass of the 
primary should be almost equal to the mass of the secondary. The clock for the explosion in this scenario 
is given by the lifetime of the secondary plus the gravitational time delay 
(Matteucci \& Tornamb\'e 1986). 

\item The C-deflagration of a Chandrasekhar mass ($\sim 1.4
M_{\odot}$) C-O WD after accretion from a non-degenerate companion
(Whelan and Iben 1973; Munari and Renzini 1992; Kenyon et
al. 1993). This model is known as the single-degenerate (SD) one. The
main problem with this scenario is the narrow range of permitted
values of the mass accretion rate in order to obtain a stable
accretion, instead of an unstable accretion with a consequent nova
explosion and mass loss. In this case, in fact, the WD never achieves
the Chandrasekhar mass. In particular, Nomoto, Thielemann \& Yokoi
(1984) found that a central carbon-deflagration of a WD results for a
high accretion rate ($\dot M > 4 \cdot 10^{-8}\, M_{\odot} \,{\rm
yr}^{-1}$) from the secondary to the primary star (the WD). They found
that $\sim 0.6 - 0.7\; M_{\odot}$ of Fe plus traces of elements from C
to Si are produced in the deflagration, well reproducing the observed
spectra. The clock for the explosion here is given by the lifetime of the secondary star.

\item A sub-Chandrasekhar C-O WD exploding by He-detonation induced by
accretion of He-rich material from a He star companion (Limongi and
Tornamb\'e 1991).

\item A more recent model by Hachisu et al. (1996; 1999) is based on the
classical scenario of Whelan and Iben (1973) (namely C-deflagration in
a WD reaching the Chandrasekhar mass after accreting material from a
star which fills its Roche lobe), but they find an important
metallicity effect. When the accretion process begins, the primary
star (WD) develops an optically thick wind which helps in stabilizing
the mass transfer process. When the metallicity is low ($[Fe/H] <
-1$), the stellar wind is too weak and the explosion cannot
occur. The clock for the explosion here is also given by the lifetime of the secondary 
star plus the chemical evolution delay time, due to the fact that the progenitors of 
Type Ia SNe do not form before the gas has attained the threshold metallicity.

\end{itemize}


\subsection{The formulation of the SNIa rate of GR83 and MR01}

In the formulation of the Type Ia rate by GR83, 
based on the Whelan and Iben (1973) model,
the explosion times correspond to the lifetimes of
stars in the mass range $0.8 - 8 M_{\odot}$. In fact, the maximum
initial mass which leads to the formation of a C-O WD is $\sim 8
M_{\odot}$, although stellar models with overshooting predict a lower
value (e.g. Marigo et al. 1996), which means that the first system,
made of two $8 M_{\odot}$ stars, explodes after $\sim 3-4 \cdot 10^7$
years from the beginning of star formation. The minimum total mass of the
binary system is assumed to be 3 M$_{\odot}$, to ensure that the WD and the
companion are large enough to allow the WD with the minimum possible mass
($\sim 0.5M_{\odot}$) to reach the Chandrasekhar
mass limit after accretion. The smallest possible secondary mass is $0.8
M_{\odot}$ and therefore the maximum explosion time is the age of the
universe. This ensures that this model is able to predict a present
time SN Ia rate for those galaxies where star formation must have
stopped several Gyr ago, such as ellipticals.

In this formalism the SNIa rate for an instantaneous starburst, in other words 
the DTD function, can be written as:

\begin{equation}
R_{\rm Ia}(t)=A \int_{M_{\rm B, inf}}^{M_{\rm B, sup}}\phi(M_{\rm B})
f\biggl({M_2(t)\over M_{\rm B}}\biggr){dM_{\rm B} \over M_B},
\end{equation}
where $M_B = M_1 + M_2$ is the total mass of the binary system, 
with $M_1$ being the primary stars 
(the originally most massive one) and $M_2$ being the secondary star,
$M_{\rm B, inf}$ and $M_{\rm B, sup}$ are the minimum and maximum
masses for the binary systems contributing at the time $t$. The
maximum value that $M_B$ can assume is called $M_{BM}$ and the minimum
$M_{Bm}$. These values (maximum and minimum mass of the binary systems
able to produce a SNIa explosion) are model-dependent. In particular,
GR83 considered that only stars with $M \leq 8 M_{\odot}$ could
develop a degenerate C-O core, thus obtaining an upper limit $M_{BM} =
16 M_{\odot}$ for the mass of the binary system. The adopted lower
limit is $M_{Bm} = 3 M_{\odot}$, as discussed previously. 
The constant $A$ represents a free parameter which indicates the fraction 
of  binary systems of the type
necessary to produce Type Ia SNe relative to all the stars in the mass 
range 3-16$M_{\odot}$.
This parameter is fixed by reproducing the present time SN Ia rate.

The extremes of the
integral (1) are functions of time and for a fixed time $t$, are:

\begin{equation}
M_{\rm B, inf} = max(2 M_2(t), M_{Bm})
\end{equation}

\begin{equation}
M_{\rm B, sup} = {1\over 2} M_{BM} + M_2(t),
\end{equation}
\noindent
where $M_{\rm B, sup} = M_{BM}$ when $M_2(t) = 8 M_{\odot}$.

We define $\mu=M_2/M_{\rm B}$ as the mass fraction of the secondary
and $f(\mu)$ is the distribution function of this ratio. Statistical
studies (e.g. Tutukov \& Yungelson 1980) indicates that mass ratios
close to one are preferred, so the formula:

\begin{equation}
f(\mu)=2^{1+\gamma}(1+\gamma)\mu^\gamma,
\end{equation}
\noindent
is commonly adopted, with $\gamma$=2 as a parameter. 

The function $\phi(M_B)$ is the initial mass function (IMF) and has the form:

\begin{equation}
\phi(M_{\rm B}) = C M_B^{-(1+x)}
\end{equation}
\noindent
where $x$ is the so-called  Salpeter (1955) index, the IMF is defined in
the mass interval 0.1-100 $M_{\odot}$, and C is the normalization 
constant (see eq. 12).The IMF that we will use will be either the Salpeter 
one with x=1.35 all over the mass range or a multi-slope IMF as suggested by 
Scalo (1986).

In order to compute the Type Ia SN rate we need to convolve the DTD function 
described above with a suitable star formation 
rate. 

In particular: 

\begin{equation}
R_{\rm Ia}(t)=A \int_{M_{\rm B, inf}}^{M_{\rm B, sup}}\phi(M_{\rm B})
\int_{\mu_{\rm min}}^{\mu_{\rm max}}f(\mu)
\psi(t-\tau_{M_2})d\mu\, dM_{\rm B},
\end{equation}

The star formation rate
in this case has to be evaluated at the time $(t-\tau_{M_2})$, with
$\tau_{M_2}$ being the lifetime of the secondary star and the clock for the explosion.
The star formation rate (SFR) $\psi(t)$
will be chosen according to the morphological type of the galaxies we will consider.

\begin{figure}
\centering
\epsfig{file=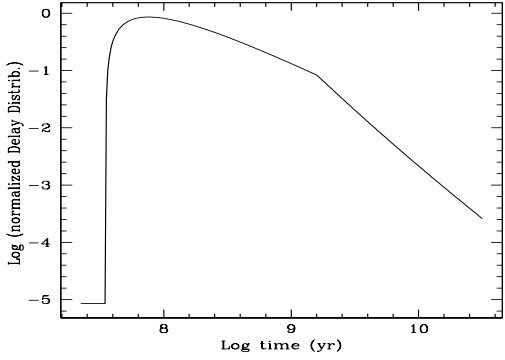,height=5cm,width=7cm}
\caption[]{The normalized time delay distribution function (DTD) as 
obtained from eq. (1) according to MR01.}
\end{figure}
In Figure 1 we show the normalized DTD from MR01. It represents the results of eq. (1), namely the Type Ia SN 
rate computed under the scenario of the single-degenerate for an instantaneous burst.The assumed IMF is the one-slope
Salpeter (1955) one.

\subsection{The new formulation of the Type Ia SN rate}

In Figure 2  we show the normalized DTD of MVP06.
We have approximated this DTD with the following expressions:

\begin{equation}
log DTD(t)=1.4 -50(log{\rm t} -7.7)^{2}
\end{equation}

for $t< 10^{7.93}$ yr, and

\begin{equation}
log DTD(t)=-0.8 - 0.9(log{\rm t} -8.7)^{2}
\end{equation}

for $t> 10^{7.93}$ yr, where the time is expressed in years.

It is worth noting that, in order to semplify the comparison, we have imposed
that the number of SNe Ia in the MR01 formulation (i.e. the integral of the 
rate in Figure 1) be the same as the number of SNe Ia  obtained from Figure 2 
with the DTD of MVP06.

This new formulation, being analytical, is easier to implement in a galactic 
chemical evolution code,
since the Type Ia SN rate for any history of star formation can be derived, by
following the formalism developed by Greggio (2005) we can write:

\begin{equation}
R_{Ia}(t)=k_{\alpha} \int^{min(t, \tau_x)}_{\tau_i}{A (t-\tau) \psi(t-\tau) 
DTD(\tau) d \tau}
\end{equation}

\begin{figure}
\centering
\epsfig{file=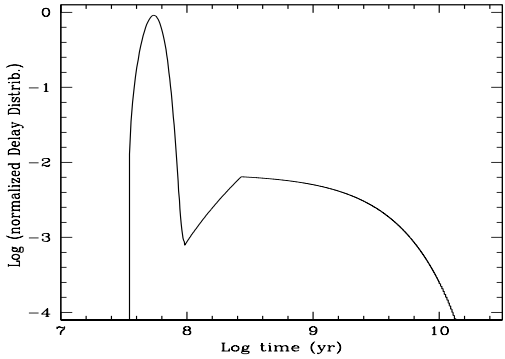,height=5cm,width=7cm}
\caption[]{The normalized time delay distribution function (DTD) as proposed 
by MVP06.}
\end{figure}

where $A(t- \tau)$ is the fraction of binary systems which give rise  to Type Ia SNe, in analogy with 
eq. (1) and in principle it can vary in time. Here we will assume it constant. It is worth noting that the fraction A in this new formulation has a different 
meaning than in the old one.
In particular, in the new formulation is the fraction of binary systems with those particular characteristics 
to give rise 
to Type Ia SNe relative to the whole range of star masses (0.1-100$M_{\odot}$) not only relative to 
the mass range 3-16$M_{\odot}$ as it is in MR01. 
The time $\tau$ is the delay time defined in the range $(\tau_i, \tau_x)$ so that:

\begin{equation}
\int^{\tau_x}_{\tau_i}{DTD( \tau) d \tau}=1
\end{equation}
where $\tau_i$ is the minimum delay time for the occurrence of Type Ia SNe, 
in other words the time at which the first SNe Ia start occurring. We assume, 
for
this new formulation of the SN rate that $\tau_i$ is the lifetime of a 
8$M_{\odot}$, while for $\tau_x$, which is the maximum delay time, 
we assume the lifetime of a  $0.8M_{\odot}$,
exactly as in the single degenerate scenario described before. 

Finally, $k_{\alpha}$ is the number of stars per unit mass in a stellar 
generation and contains the IMF.

In particular:
\begin{equation}
k_{\alpha}= \int^{m_U}_{m_L}{\phi(m)dm}
\end{equation}

where $m_L=0.1 M_{\odot}$ and $m_U=100 M_{\odot}$ and define the whole range 
of existence of the stars.

The normalization condition for the IMF is the usual one:
\begin{equation}
\int^{m_U}_{m_L}m \phi(m)dm =1
\end{equation}

\subsection{Which systems could explain the DTD of MVP06?}

A possible justification for this strongly bimodal DTD of MVP06 can be found in the framework of the single degenerate model for 
the progenitors of Type Ia SNe. In fact, such a DTD can be found if one assumes that the the function 
describing the distribution of mass 
ratios inside the binary systems, $f(\mu)$, as defined before, is a multi- slope function. In particular,
a slope  $\mu = 2.0$ should be assumed  for systems with stars in the mass range 5-8 $M_{\odot}$, 
whereas a negative slope
($\mu \sim -0.8 / -0.9$) should be adopted for masses lower than 5$M_{\odot}$.
This choice means that in the range  5-8 $M_{\odot}$ are preferred the systems where $M_1 \sim M_2$, whereas 
for lower mass progenitors are 
favored systems where $M_1 >> M_2$.
However, this expression for $f(\mu)$
should be compared with observational estimates of the mass ratio distribution function.
In general, there is not yet an agreement on a universal form for the $f(\mu)$ or the $f(q)$ 
($q= {M_2 \over M_1}$) function,
due to the fact that most stellar samples and observational techniques
are affected by observational biases.
As already mentioned, Tutukov \& Yungelson (1980) suggested a one slope $f(\mu)$ 
function with $\gamma=2.0$,
whereas Duquennoy and Mayor (1991), from a sample of solar type stars, suggested a 
one-slope function with  a value of $\gamma$=-0.35.

More recently, Kouwenhoven et al. (2005) from A and B type stars in the Scorpius-Centaurus association 
found a $f(q)=q^{-0.33}$, while  Shatsky \& Tokovinin (2002), by studying B 
stars always in the Scorpius-Centaurus association, suggested $f(q)=q^{-0.55}$.
Hogeveen (1992) studied the mass ratio distribution of spectroscopic binaries 
and obtained $f(q)\propto q^{-2}$ for $q>q_o$
with $q_o=0.3$  whereas for $q <q_o$ he found a flat distribution. He also estimated that a 
fraction of 19-45\% of 
the stars in the solar neighbourhood are spectroscopic binaries.
Finally, for stars with masses in the range 10-20$M_{\odot}$, Pinsonneault \& Stanek (2006) found evidence for the existence of two distinct classes of binary 
systems, each one contributing  to about half of the systems: one being 
characterized by two stars of similar mass and the other by a flat distribution of mass ratios.

In conclusion, the observational situation is still too uncertain to draw firm conclusions and it is better to use 
the analytical formulation for the DTD, as described before, without worrying about the precise SNIa progenitor 
model.

Now we can proceed in computing the chemical evolution of galaxies of different morphological type by means 
of this new formulation of 
the SNIa rate, as it will be described in the next sections.

\section{Models of chemical evolution} 

We describe here the general equations for the chemical evolution of galaxies 
and the 
assumptions common to all models.
The different morphological types are then identified by the different star 
formation histories 
(namely by the SFR, $\psi(t)$; and by the IMF; $\phi(m)$) and by different 
infall /outflow rates.

The time-evolution of the fractional mass of the element $i$ 
in the gas within a galaxy, $G_{i}$, is described by the basic 
equation:

\begin{equation}
\dot{G_{i}}=-\psi(t)X_{i}(t) + R_{i}(t) + (\dot{G_{i}})_{inf} -
(\dot{G_{i}})_{out}
\end{equation}

where $G_{i}(t)=M_{g}(t)X_{i}(t)/M_{tot}$ is the gas mass in 
the form of an element $i$ normalized to a total fixed mass 
$M_{tot}$ and $G(t)= M_{g}(t)/M_{tot}$ is the total fractional 
mass of gas present in the galaxy at the time t.
The quantity $X_{i}(t)=G_{i}(t)/G(t)$ represents the 
abundance by mass of an element $i$, with
the summation over all elements in the gas mixture being equal 
to unity. $\psi(t)$ is the fractional amount of gas turning into 
stars per unit time, namely the SFR. 
$R_{i}(t)$ represents the returned fraction of matter in the 
form of an element $i$ that the stars eject into the ISM through 
stellar winds and supernova explosions; this term contains all 
the prescriptions concerning the stellar yields and 
the supernova progenitor models. 

The nucleosynthesis prescriptions are common to all models and are taken from: 
Woosley \& Weaver (1995) yields 
for massive stars (those relative to the solar chemical composition), van den 
Hoeck $\&$ Groenewegen 
(1997) for low and intermediate mass stars 
($0.8 \le M/M_{\odot} \le 8$) and Nomoto et al. 
(1997) for Type Ia SNe.

The assumed stellar lifetimes for all galaxies are those suggested by 
Padovani \& Matteucci (1993).

The two terms 
$(\dot{G_{i}})_{inf}$ and  $(\dot{G_{i}})_{out}$ account for 
the infall of external gas and for galactic winds, respectively.
The presence of infall and winds varies for galaxies of different 
morphological type. Infall is present in all galaxy models but with different 
timescales, whereas the wind is present in ellipticals and dwarfs but absent 
in spirals.

The prescription adopted for the star formation history is the
main feature which characterizes a particular morphological 
galactic type.

In its simplest form the SFR, $\psi(t)$, in our models is a 
Schmidt (1959) law expressed as:

\begin{equation}
\psi(t) = \nu G^{k}(t)
\end{equation}

The quantity $\nu$ is the efficiency of star formation, 
namely the inverse of the typical time-scale for star formation,
and is expressed in $Gyr^{-1}$.

The rate of gas infall is defined as:
\begin{equation}
(\dot G_{i})_{inf}\,=\,Ce^{-t/ \tau}
\end{equation}

with C being a suitable constant and $\tau$ the infall timescale.

The rate of gas loss via galactic winds for each element {\it i} is
assumed to be proportional to the star formation rate at the 
time {\it t}:

\begin{equation}
 \dot G_{iw}\,=\,w_{i} \, \psi(t)
\end{equation}

where $w_{i}$ is a 
free parameter describing the efficiency of the galactic
wind. 

In all models the instantaneous recycling approximation is relaxed and the 
stellar lifetimes are taken into account.

\subsection{The Milky Way} 
The role played by SNe of different Type(II, Ia) in the chemical 
evolution of the Galaxy has been computed by several authors 
(Matteucci \& Greggio, 
1986; Yoshii et al. 1996; Tsujimoto et al. 1995; Chiappini et al. 1997; 
Kobayashi et al. 1998; Boissier \& Prantzos 1999).
Here we refer to the model of Chiappini et al. (1997), the so-called two-infall model for the 
evolution of the Milky Way. A thorough description of this model can be found in Chiappini et al. 
(1997; 2001; 2003) and Fran\c cois et al. (2004) and we address the reader to these paper for 
details.
It is assumed that the stellar halo formed on a relatively short 
timescale (1-2 Gyr) by means of a first infall episode, whereas the disk 
formed much more slowly mainly out of 
extragalactic gas thanks to a second infall episode. The timescale for the 
disk formation is assumed to increase 
with the galactocentric distance ($\tau= 7$ Gyr at the solar circle), 
thus producing an ``inside-out'' scenario 
for the disk formation.The Galactic disk is divided in several rings 2Kpc 
wide without exchange of matter between them.  

The model can follow in detail the evolution of several chemical 
elements including H, D, He, C,N,O, $\alpha$-elements, Fe and Fe-peak elements, 
s- and r-process elements.
The IMF is assumed to be constant in space and time and is that of Scalo(1986).

The star formation rate adopted for the Milky Way  is a function of both 
surface gas density and total surface mass density.
Such a SFR is proportional to a power $k = 1.5$ of the
surface gas density and to a power $h = 0.5$ of the total surface mass
density. This formulation of the SFR (details can be found in Chiappini et al. 1997;2001) 
takes into account the feedback
mechanism between stars and gas regulating star formation and is
supported by observations (e.g. Dopita and Ryder 1994).In the Milky Way model 
we also assume a surface density threshold below which the SFR stops, 
according to Kennicutt (1989;1998).As a consequence of
this, the star formation rate goes to zero every time that the gas
density decreases below the threshold ($\sim 7
M_\odot\,pc^{-2}$).
The efficiency of SF is $\nu=0.1 Gyr^{-1}$ during the disk 
and 2$Gyr^{-1}$ in the halo phase. 
This  model reproduces the majority of the features
of the solar vicinity and the whole disk (see Chiappini et al, 1997;2001).

\subsection{Elliptical galaxies}

For the chemical evolution of ellipticals we adopt the model of Pipino \& 
Matteucci (2004) where we address the reader for details.

Here we recall the main assumptions:
ellipticals form by means of a fast collapse of pristine gas where star 
formation occurs at a very high rate (starburst-like regime), and after a 
timescale, varying with the galactic mass, each galaxy develops a galactic winds
due to the energy deposited by SNe into the interstellar medium (ISM).After 
the development of this wind no star formation is assumed to take place.
The SN feedback is taken into account together with the cooling of SN 
remnants and the development of galactic winds is calculated in a 
self-consistent way. Massive but diffuse haloes of dark matter around these 
galaxies are considered.
We assume that the efficiency of star formation rate is higher in more 
massive objects which evolve faster than less massive ones (inverse-wind 
scenario, Matteucci, 1994, otherwise called ``downsizing'', Cowie et al. 1996).
 The Salpeter (1955) IMF constant in space and time is adopted.
In the SFR expression (eq. 14) we assume k=1 and $\nu =1-20 Gyr^{-1}$, going from $10^{9}$ to
$10^{12}M_{\odot}$ of luminous mass. 
The infall rate is quite fast with a typical $\tau=0.5$Gyr.
Here we will consider the model corresponding to a typical  elliptical of $10^{11}M_{\odot}$
of luminous mass.

\subsection{Dwarf irregulars}
For the dwarf irregulars we assume a
one zone model with instantaneous and complete mixing of gas inside
this zone (see Lanfranchi \& Matteucci, 2003). The development of a 
galactic wind is computed similarly to what is 
done for ellipticals except that in these small objects the wind does not 
stop the star formation. The star formation takes place in short bursts of 
duration not longer than 200 Myr. The star formation rate is again a 
Schmidt law (eq. 14) with $k=1$ and $\nu=(0.1-0.9) Gyr^{-1}$.
The infall timescale is also quite short ($\tau \sim 0.3-0.5Gyr$)
and the IMF is the Salpeter one.

\section{Model Results}
In this section we will show a comparison between results 
obtained by means of the just described models
for 
different galaxies with the old formulation of the SNIa rate (GR83; 
MR01) and those obtained with the new rate as described in Section 2.

\subsection{The Milky Way}

\begin{figure}
\centering
\epsfig{file=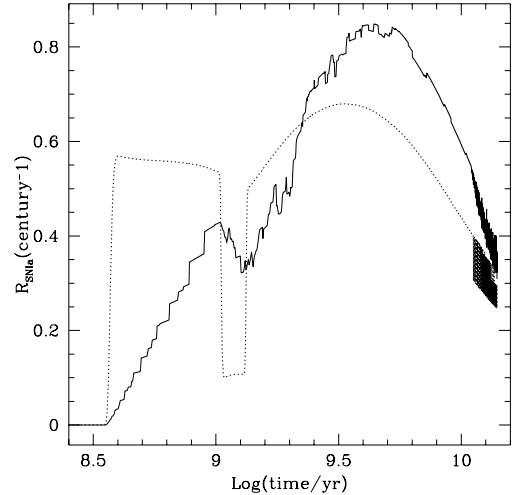,height=7cm,width=7cm}
\caption[]{The Type Ia SN rate (expressed in SN $Gyr^{-1}$) as a 
function of time in the Milky Way as obtained for 
the old prescription of the rate (solid curve) and for the new one (dotted curve).The SFR is 
the same but the DTDs 
are different, as described in the text.The parameter A=0.09 for the MR01 
formulation and A=0.0025 for the new one (see text). 
In both cases the IMF is the Scalo (1986) one, which is appropriate for the
disk of the Milky Way.}
\end{figure}

In Figure 3 we report the model results for the Type Ia SN rate as a function of cosmic time 
in the cases of the old rate (eq. 5) 
and the new rate (eq. 9).The oscillating 
behaviour at late times, in both cases,  is due to the adoption 
of a threshold gas density in the SFR. 
The parameter $A$, as previously defined, is $A=0.09$ for the old formulation 
of the SN Ia rate and for the assumed stellar lifetimes. In fact, if one adopts different stellar lifetimes such as Maeder \& Meynet (1989), the value of the parameter is $A=0.05$ (see Romano et al. 2005 for a detailed discussion).The parameter $A$ indicates the fraction of binary systems producing 
Type Ia SNe in the mass range 3-16$M_{\odot}$ relative to all the stars in 
this same mass range, and is fixed by reproducing the present time observed 
Type Ia rate in the Milky Way. The parameter $A$ depends on the assumed stellar lifetimes and IMF.
When the new formulation is adopted, this parameter assumes a different 
meaning, being the fraction of binary systems originating Type Ia SNe relative 
to whole mass range where the IMF is defined.
In particular, for the new rate we have to assume a much lower $A=0.0025$.
One can immediately notice in Figure 3 that 
the integrals under the two curves are roughly the same whereas the shapes 
of the curves are quite different.
In particular, the old Type Ia SN rate shows a minor peak at around 1Gyr, 
then a minimum is reached at $\sim$
1.4 Gyr and this
is the consequence of the gap in the SFR predicted by the two-infall model 
between the halo-thick-disk and the 
thin-disk phases. In the new rate the peak at early times is much more 
pronounced and extended, reflecting the peak in 
the DTD of MVP06 (prompt Type Ia SNe), then the minimum due to the gap in 
the SFR is more pronounced than before and it starts at 1Gyr.
The lower values of the new rates after the gap are again a consequence of 
the assumed DTD.  
The value of A=0.0025 for the new rate is fixed by obtaining a present time 
rate as close as possible to the 
observed one ($\sim$ 0.3 SNe$(100 yr)^{-1}$, Cappellaro et al. 1999) without 
overestimating the number of SNeIa in the past.In particular, with the new 
formulation we obtain a present time Type Ia SN rate of 
0.25  SNe $(100 yr)^{-1}$, well in agreement with the observed rate inside the 
observational errors.

At this point,  we should check the effects of the new rate on the 
age-metallicity relation and on [O/Fe] versus [Fe/H] relation, shown in 
Figures 4 and 5, respectively.
In Figure 4 we show the plot of the age-metallicity relation as obtained
for the solar vicinity. Two cases are shown: the results with the old and new Type Ia SN rates. 
It is worth noting that in the old formulation of Chiappini et al. (1997) it was assumed that each Type Ia SN was 
producing on 
average $\sim 0.4 M_{\odot}$ of Fe. This was done to take into account the fact that the data  
indicates that not all Type Ia SNe produce $\sim 0.6-0.7 M_{\odot}$ of Fe but that there is a certain 
spread. This lower Fe mass was preventing, with the old rate, a too high solar Fe abundance.
With this prescription for Fe in SNeIa, the new rate produces a slightly low solar Fe, as a consequence of the 
differences between the old and new rate. 
Therefore, we adopted here, for the sake of homogeneity with the other galactic models presented here, 
the canonical 0.7$M_{\odot}$ of Fe per SN (Nomoto et al. 1997) which, by the 
way, represents an even better average for the Fe mass. In fact, 
the spread found in the Fe among Type Ia SNe is as large 
as 0.1-1.1$M_{\odot}$ (Cappellaro et al. 1997).By doing that, the age- metallicity relation, 
obtained by means of the new rate, is very similar to the old one. 
However, the solar Fe abundance obtained with the new rate and this 
prescription 
is still lower than the old one and in better agreement with the latest solar abundance determination by 
Asplund et al. (2005) (see Figure 4, where the theoretical solar abundance corresponds to 
the value at a time 9.5 Gyr, having assumed a galactic age of 14 Gyr).

\begin{figure}
\centering
\epsfig{file=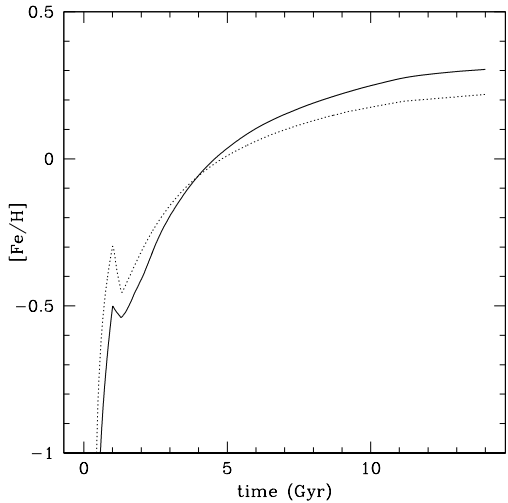,height=7cm,width=7cm}
\caption[]{The computed age-metallicity relation in the solar neighbourhood
 for the old formulation of the DTD (MR01) (continuous line) and the new one 
(dotted line) with an average Fe per Type Ia SN of 
$0.7M_{\odot}$. The Fe abundances are normalized to the 
solar values of Asplund et al. (2005).}  

\end{figure}

In Figure 5 we show the same models of Figure 4 showing the [O/Fe]
versus [Fe/H] relation. In this case, the [O/Fe] ratio obtained by means of 
the new rate is similar to the one in the case with the old rate for very low [Fe/H], but for 
[Fe/H]$> -2.0$ it decreases faster. The maximum difference between the two curves occurs at [Fe/H]$\sim$ -1.0 dex and is 
roughly 0.2 dex. 
This is the consequence of the 
larger number of Type Ia SNe at early times obtained with the DTD of MVP06 which lowers the [O/Fe] ratio 
at a lower [Fe/H]. 
In Figure 5 we show also the data compiled by Fran\c cois et al. (2004).
The fit to the data is not the best and the reason is that here, for purposes of homogeneity with the other 
galactic models, we do not adopt the empirical yields suggested by Fran\c cois et al. (2004) to best fit the data. 
In particular, the main 
difference is in the oxygen yields which here are those from Woosley \& Weaver (1995) for solar chemical composition, 
whereas Fran\c cois et al. (2004) adopted those as functions of the metallicity for the best fit. The best model of 
Fran\c cois et al. (2004) is shown for 
comparison in Figure 6. This model differs from the others in the oxygen 
yields and in the stellar lifetimes which are those of Maeder \& Meynet (1989).
In Figure 6, we also plot the best model of Fran\c cois et al. when the MVP06 
rate is assumed. As for the models of Figure 5 we can see that the main 
difference is the fast decrease of the [O/Fe] ratio for [Fe/H]$>$ -2.0 dex.
As it is evident from an inspection of Figures 5 and 6, both formulations of the Type Ia SN rate can fit the data, although the old one 
produces a better agreement with the data than the new one. It is worth noting that in the DTD of MR01 the percentage of systems exploding within $10^{8}$ years is $\sim 13 \%$, whereas in the new DTD is $\sim 50 \%$. Our results indicate that perhaps this fraction is too high. On the other hand, the results of MVP06 indicates that this percentage can be as low as 35-40 \% and still produce a good agreement with the radio data.  

\begin{figure}
\centering
\epsfig{file=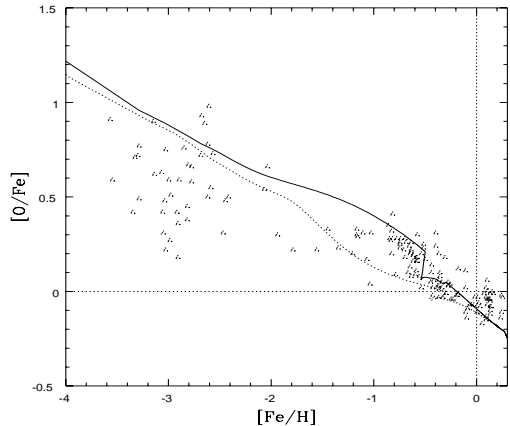,height=6cm,width=7cm}
\caption[]{The computed  [O/Fe]
versus [Fe/H] relations for the same cases of Figure 4 (dotted line, MVP06 formulation, continuous line, MR01 formulation). 
The abundances of all models are normalized to the latest solar 
values of Asplund et al. (2005). The data are from the compilation of 
Fran\c cois et al. (2004). The horizontal and vertical lines represent the 
solar values.}  

\end{figure}

\begin{figure}
\centering
\epsfig{file=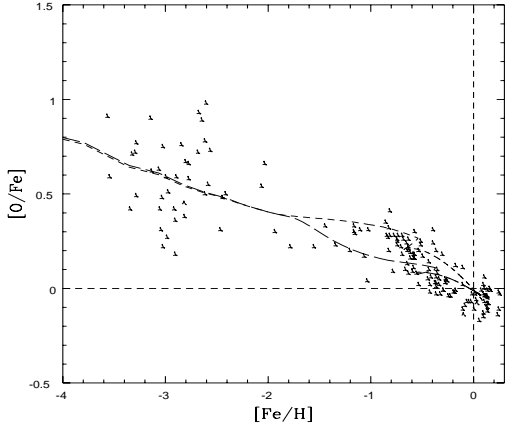,height=6cm,width=7cm}
\caption[]{The computed [O/Fe]
versus [Fe/H] relations for the best model of Fran\c cois et al. (2004) 
(short-dashed line); this particular model differs from the models of Figure 5 
for the yields of oxygen in massive stars but the difference does not produce 
noticeable differences.
More important is the difference in the stellar lifetimes which are those of
Maeder \& Meynet (1989), whereas in the models of Figure 5 and in all the 
models of this paper are from Padovani \& Matteucci (1993). Here we show also 
the results of the best model of Fran\c cois et al. (2004) 
with the MVP06 
rate (long-dashed curve). Also in this case the mass of Fe from each Type Ia 
SN was assumed to be 0.70$M_{\odot}$ in analogy of what we did for the dotted 
model of Fig.5 (see text). The abundances of all models are 
normalized to the latest solar values of Asplund et al. (2005). 
The data are from the compilation of Fran\c cois et al. (2004).
The horizontal and vertical lines represent the solar values.}  

\end{figure}

\subsection{Ellipticals}

In Figure 7 we show the computed Type Ia SN rates for a typical 
elliptical of $10^{11}M_{\odot}$ of initial luminous mass for the old and new 
formulations. The final stellar mass after the galactic wind, for this 
model, is $M_{*}=3.5 \cdot 10^{10}M_{\odot}$ and the blue luminosity is 
$L_{B}=4 \cdot 10^{9} L_{B_{\odot}}$.  
Note the very high values of the rates at early times, due to the 
very high star formation rate during the formation of the spheroids 
($\sim 1000 M_{\odot} yr^{-1}$), obtained by our model.

\begin{figure}
\centering
\epsfig{file=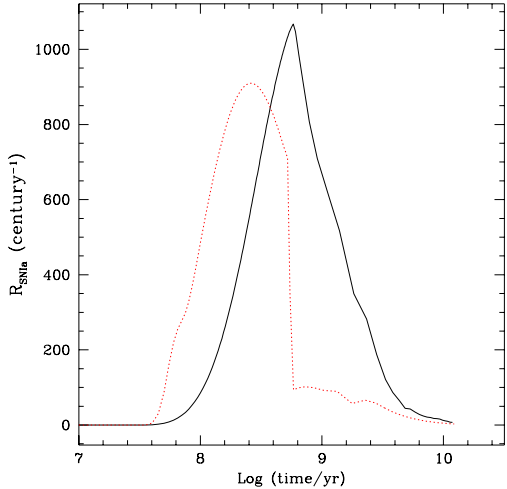,height=7cm,width=7cm}
\caption[]{The computed Type Ia SN rate in the old and new formulation 
for 
a typical elliptical galaxy of $10^{11} M_{\odot}$ of initial luminous mass. 
The solid line is the SN Ia rate as obtained with the old formulation 
and 
the A=0.18.The dotted  line represents 
the new rate computed with the DTD of MVP06 and A=0.0045.In both cases the 
adopted IMF is the Salpeter one.}

\end{figure}

In the case of ellipticals, as we can see from Figure 7, it is possible to 
obtain almost the same present time 
Type Ia SN rate as in the old formulation ($\sim$ 0.04 SNe $(100 yr)^{-1}$ in 
agreement with Cappellaro et al. (1999) observed rate), only by changing the 
value of the A parameter, which is A=0.18 in the
old formulation and A=0.0045 in the new one.

The new rate presents a 
higher number of SNe Ia at very early times but at late times coincides 
exactly with old one.
In the case of an elliptical galaxy, in fact, the SFR proceeds like in a 
burst and for the majority of the 
cosmic time the galaxy is evolving passively. This is why there is not a great 
difference between the two rates at late times.

In Figure 8 we present the  [O/Fe] versus [Fe/H] for the two cases: 
the old model of Pipino \& Matteucci (2004) and the model with the 
new Type Ia rate.
In this figure we can see the effect of the new rate on the [O/Fe] vs. [Fe/H] 
relation computed for the gas in a typical elliptical galaxy. 
In analogy with the Milky Way, the differences among the models with the new 
and old rate consist in a faster decrease of the [O/Fe] ratio for 
[Fe/H]$> -1.0$ 
in the model with the new rate, due to the peak in the SNeIa 
obtained with the 
MVP06 DTD.The fact that in ellipticals the difference in the [O/Fe] ratio 
appears for [Fe/H] $>$ -1.0 dex, whereas in the Milky Way it appears for 
[Fe/H]$>$ -2.0 dex is due to the more efficient star formation in ellipticals
which allows the gas to reach a high metallicity in a very short time.

\begin{figure}
\centering
\epsfig{file=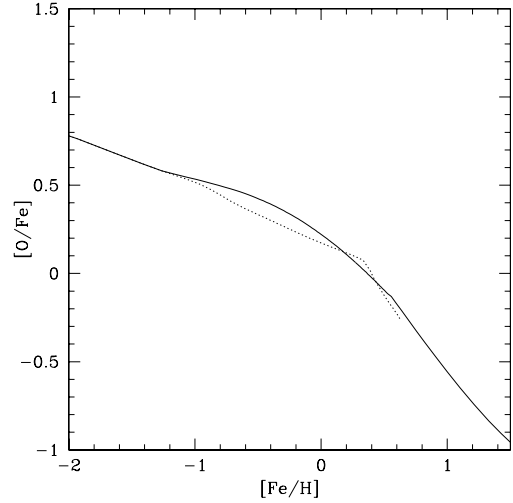,height=7cm,width=7cm}
\caption[]{The computed [O/Fe] vs. [Fe/H] in the framework of the
old (MR01) and new (MVP06) formulation
for the Type Ia SN rate for a typical elliptical of $10^{11}M_{\odot}$ of 
luminous mass.The solid line is the relation obtained with 
the old formulation 
of MR01 with A=0.18 for a Salpeter (1955) IMF(see Pipino \& Matteucci 2004), 
whereas the dotted one is the relation obtained with with the new 
formulation, the Salpeter IMF and A=0.0045.}
\end{figure}

Again, like for the case of the Milky Way we can conclude that the new 
formulation of the Type Ia SN rate leads to a higher Type Ia SN rate at early 
times and to a faster decrease of 
this rate at late times, but it does not affect substantially the galactic 
chemical evolution.It is worth noting that in 
the Pipino \& Matteucci (2004) model the Fe mass produced by each SN Ia was 
assumed to be 0.7$M_{\odot}$ and we did not change this value in the new 
formulation. The only noticeable difference  between the new and old model 
is the maximum [Fe/H] reached in the gas. In the new model is [Fe/H]=+0.6 dex, 
whereas in the old one is [Fe/H] $>$ +1.0 dex. This fact does not have 
implications for the [Fe/H] in the stars in ellipticals but only for the 
amount of Fe ejected by ellipticals in the intracluster medium (ICM), 
as we will see in paragraph 5.4.

\subsection{Dwarf irregulars}
In Figure 9 we show the Type Ia SN rate for a typical dwarf starbursting 
irregular galaxy
of luminous mass $10^{8} M_{\odot}$. The old and new rates are compared for an 
euristic case with three starbursts of 0.2 Gyr 
duration and occurring at 2, 6 and 10 Gyr, respectively. The star formation 
efficiency is chosen to be $\nu=0.1 Gyr^{-1}$ in each burst.

\begin{figure}
\centering
\epsfig{file=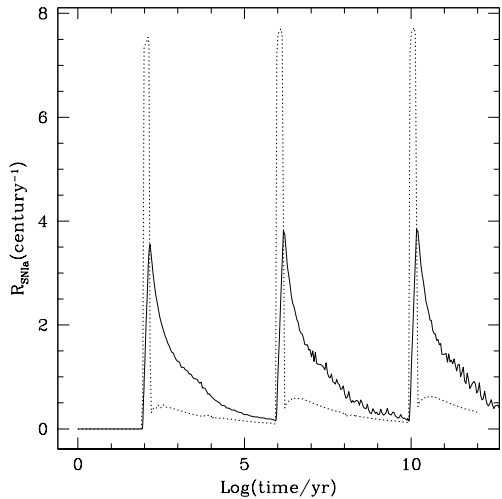,height=7cm,width=7cm}
\caption[]{The computed Type Ia SN rate (multiplied by $10^{5}$) 
as a function of cosmic time for a 
dwarf irregular galaxy with 3 bursts of star formation and efficiency of 
star formation $\nu=0.1 Gyr ^{-1}$ with the new 
(dotted line) and old (continuous line) formulation. In the old formulation 
A=0.09, whereas in the new one is A=0.0037.}

\end{figure}

As one can see from Figure 9, the main difference between the new and the old 
formulation of the Type Ia SN rate for a starburst regime consists in a larger amount of SNe Ia at the beginning of each starburst and in a lower amount of SNe in the interburst periods. The present time Type Ia SN rates in the two cases are similar and this is obtained by adopting A=0.0037 for the MVP06 case and A=0.09 for the MR01 case. The present time rates are: $3.4 \cdot 10^{-6}$ SNe $(100 yr)^{-1}$ for the MVP06 case and  $2.8 \cdot 10^{-6}$ SNe $(100 yr)^{-1}$ for the MR01 case. These are very low values because we considered an euristic  case with very few bursts and very low star formation efficiency.

\begin{figure}
\centering
\epsfig{file=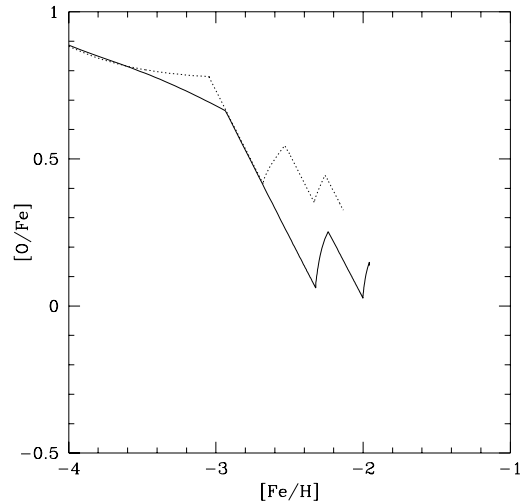,height=7cm,width=7cm}
\caption[]{The computed [O/Fe] vs. [Fe/H] for a 
dwarf irregular galaxy with 3 bursts of star formation. The parameters of the models are the same as 
in Figure 9. The continuous line represents the model with the old 
formulation of the 
Type Ia SN rate of MR01 whereas the  dotted line represents the model with the new rate (MVP06).}

\end{figure}

Finally, in Figure 10 we show the calculated [O/Fe] versus [Fe/H] for the new and old Type Ia SN rates. 
As one can see the behaviour of the [O/Fe] ratio is similar in the two cases with the saw-tooth 
behaviour typical of the star burst regime of star formation, but the case with the new rate produces 
higher [O/Fe] ratios (by $\sim 0.2$ dex) relative to the case with the old rate, especially after the first burst 
episode.The reason is that there are much less Type Ia SNe during the interburst phases than in the old formulation, and
this contributes to keep higher the [O/Fe] ratio.
Therefore, with the new rate of Type Ia SNe dwarf starburst galaxies (with very few bursts) are 
found to have higher [$\alpha$/Fe] ratios than with the old models.
For an increasing number of bursts mimicking a continuous star formation, the differences between the [O/Fe] ratios tend to disappear.
As in all the previous cases, also here the 
final [Fe/H] value rached in the gas is slightly lower for the new formulation 
of the Type Ia SN rate.

\subsection{Chemical enrichment of the ICM}
In this section we explore the effects of the new Type Ia SN rate on the chemical enrichment of the 
intracluster medium (ICM). Pipino et al.(2002) computed the contributions of the elliptical and S0 galaxies 
to the Fe abundance in clusters. Their method was based on the
first paper of this series by Matteucci \& Vettolani (1988) where the contribution to Fe and other 
heavy elements was, for the 
first time, computed by integrating the contribution from each galaxy on a Schechter (1976) luminosity function.
The main assumptions are the contributions from each galaxy
that only E and S0 galaxies contribute to the 
chemical enrichment of the ICM and that all the Fe produced by Type Ia SNe is lost into the ICM, either by galactic winds or ram pressure stripping. The main results of Matteucci \& Vettolani (1988) were:
the Fe content of both rich and poor clusters can be well reproduced whereas the contribution to the total cluster 
gas by galaxies is small, 
thus requiring that most of the gas has a primordial origin. This was confirmed by the fact that the estimated ratio 
between the mass of the ICM and the mass in galaxies in the Coma cluster is $M_{ICM}/M_{Gal}=5.45$ (for $h=70$, White et al. 1993). After this first paper, other studies concluded that the Type Ia SN rate should have been much higher in the past to explain the Fe in clusters (Ciotti et al. 1991; Renzini, 2004).
Pipino et al. (2002) confirmed these results and calculated in detail when the Fe is ejected into the ICM by stellar winds, 
as well as considered the temporal evolution of the cluster luminosity function. 
They adopted models for 
elliptical galaxies very similar to that described before. In Figure 11 we show the abundance of Fe relative to the solar one  produced by galaxies and
ejected into the ICM of a rich cluster as a function of redshift, both in the case of the MR01 rate for Type Ia SNe and for the new rate. The chemical evolution model for ellipticals is that described in sects. 3.2 and 4.2. As expected, 
the difference in the behaviour of the temporal growth of the Fe abundance in the two cases is negligible. We find only a slightly lower Fe mass for the new rate. The reason for this can  be found in the fact that with the new Type 
Ia SN rate, a larger Fe fraction is locked in stars relative to the old rate and therefore the amount of Fe which is 
ejected into the ICM is lower.
In particular, the Fe mass in the ICM in the case of the new Type Ia SN rate is $\sim 30 \%$ lower than in the case 
with the MR01 rate. 
In order to estimate the Fe abundance in the ICM in Figure 11 we have divided the total mass of Fe present in the ICM at any time by 
the total mass of gas in the cluster. For the case of a rich cluster like Coma, the amount of gas is 
$M_{ICM}=1.5 \cdot 10^{14} M_{\odot}$. In conclusion, Figure 11 shows that in both cases the Fe in the ICM is not strongly 
evolving as a function of time (see also Scannapieco \& Bildsten, 2005), and that galactic models with the new rate produce tend to produce too little Fe in the ICM. This again suggests that less 
Type Ia SNe at early times than in the DTD of MVP06 would improve the 
agreement with observations, as we concluded for the Milky Way. However, 
it is worth noting that the comparison between data and models in Figure 11 is 
only indicative since the models represent the mass- weighted Fe abundance, whereas the data represent the emission-weighted Fe abundances.These latter favor the central cluster regions where the Fe abundance is higher.In any case, the increasing Fe trend between redshift z=1 and z=0 seems to be confirmed by more recent data by Balestra et al. (2006). If this effect is real, then we should find
some other Fe sources in the ICM acting at relatively recent times.
 
\begin{figure}
\centering
\epsfig{file=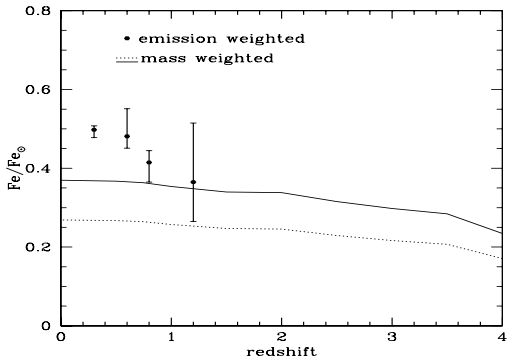,height=5cm,width=7cm}
\caption[]{The calculated $Fe/Fe_{\odot}$ in the ICM of a rich cluster 
(e.g. Coma) as a consequence of Fe mass loss by E and S0 galaxies, as a 
function of redshift.The adopted solar abundance, both in the models and 
data, is the one by Asplund et al. (2005).
The model Universe assumed here implies $\Omega_m=0.3$ and 
$\Omega_{\Lambda}=0.7$, and redshift of galaxy formation $z_f=5.0$. 
The data (emission-wighted) are from Tozzi et al. (2003) and refer to rich galaxy clusters.}

\end{figure}
\section{Summary}
In this paper we have tested a new formulation for the Type Ia SN rate in galaxies of different 
morphological type. The new formulation of the SN Ia rate is based on an empirical rate derived by 
MVP06 in order to reproduce various observational constraints relative to Type Ia SN rates and in 
particular the Type Ia 
rates  versus radio flux in radio galaxies.
It is suggested that $\sim 50\%$ of all SNe Ia should explode soon after the beginning of star 
formation and should therefore originate from stars with $M> 5M_{\odot}$, 
whereas the other half should originate from less massive progenitors.
In the framework of the single degenerate model, such a 
bimodal distribution of the Type Ia SNe could be explained if the binary systems with a unitary mass 
ratio ($M_1/M_2=1$) are 
favored in the mass range 5-8$M_{\odot}$,  
whereas for masses $< 5M_{\odot}$  the favored systems should have the mass 
of the primary much larger than the mass of the secondary.
We included  the suggested new rate in detailed models of chemical 
evolution of galaxies of 
different morphological type: the Milky Way, ellipticals and dwarf 
starburst galaxies.
In general, we found an earlier and broader peak for the Type Ia SNe in 
any type of galaxy. 
It is worth noting that a  DTD function with systems exploding after 30-40 Myr
had already been adopted by GR83 and MR01, the only difference with the 
new DTD being the lower number of 
systems exploding at early times. In particular, in these formulations the 
fraction, relative to the total number,
of Type Ia SNe exploding inside the first 0.1 Gyr since the beginning of star 
formation is $\sim 13\%$, to be compared with $\sim 50\%$ of MVP06.

We find that a 
spiral galaxy like the Milky Way should have a peak in the Type Ia rate before 
1 Gyr from the 
beginning of star formation. This implies that Type Ia SNe can pollute the ISM already during the 
halo phase, thus producing a faster decrease of the [O/Fe] ratio versus [Fe/H,] marginally in agreement  with observational data.
To this purpose, we should not forget that the MVP06 rate is an empirical one and the fraction of fast Type Ia SNe 
can be as low as 35-40\% and still produce a good fit to the radio data.
Our results suggest that less than 50\% prompt Type Ia SNe would fit better the [O/Fe] vs [Fe/H]relation in the Milky Way, in agreement with all the previous results (Matteucci \& Greggio, 1986; Chiappini et al., 2001; Fran\c cois et al. 2004.).

For ellipticals the differences are less noticeable since we assume that elliptical galaxies, 
especially the massive ones, have evolved very fast and at a very high redshift. 
The differences produced in the [O/Fe] ratio in this case are negligible.

The same situation is true for dwarf starbursting objects where the main effect is a larger number 
of Type Ia SNe at the early stages of each burst and less SNe in the interburst phases, during which only Fe is produced. In this 
case, the effect of the new rate on the computed [O/Fe] versus [Fe/H] relation is larger, due to the 
high number of fast SNe Ia in each burst and the reduced number of them in the interburst phases.
As a consequence of this, the [$\alpha$/Fe] ratios in galaxies with a small 
number of bursts tend to be always oversolar.

We have also checked the effect of the new Type Ia SN rate on the chemical 
enrichment of the ICM: the behaviour 
of the Fe abundance as a function of cosmic time does not change but less Fe 
mass is ejected in total into the ICM 
relative to previous models (roughly 30\% less). This is due to the fact that 
the larger amount of Fe produced at early times is going to be locked into 
stars, thus lowering the fraction of Fe which will be ejected into the ICM. This again suggests a fraction of prompty Type Ia SNe lower than 50\%.

\section*{Acknowledgments}
We would like to thank L.B. Yungelson and M.B.M. Kouwenhoven for sending us 
information on the latest work on observed frequencies of binary systems.
F. Matteucci especially thanks the Space Telescope Institute where this work 
was started during her partecipation in the Caroline Herschel Distinguished 
Lectures Program.

\bsp

\end{document}